\documentclass[twocolumn,secnumarabic,amssymb,superscriptaddress, nobibnotes, nofootinbib,aps,prd]{revtex4-1}

\usepackage{mathrsfs}
\usepackage{physics}
\usepackage{tikz}
\usepackage[caption=false]{subfig}
\usepackage[colorlinks=true,linkcolor=blue,citecolor=blue,urlcolor=blue]{hyperref}
\usepackage{cleveref}
\usepackage{amsmath,amssymb,amsfonts,amsthm}
\usepackage{graphicx}
\usepackage{ragged2e}
\usepackage{float}
\usepackage{bm}

\def\ra{\rangle}
\def\la{\langle}

\def\no{\nonumber}
\def\bea{\begin{eqnarray}}
\def\eea{\end{eqnarray}}
\def\be{\begin{equation}}
\def\ee{\end{equation}}

\usepackage{filecontents}

\begin{document}

\title{Thermality of the Unruh effect with intermediate statistics}%
\author{Jun Feng}%
\email{Corresponding author\\
j.feng@xjtu.edu.cn}
\affiliation{School of Physics, Xi'an Jiaotong University, Xi'an, Shaanxi 710049, China}
\affiliation{Institute of Theoretical Physics, Xi'an Jiaotong University, Xi'an, Shaanxi 710049, China}

\author{Jing-Jun Zhang}
\affiliation{School of Physics, Xi'an Jiaotong University, Xi'an, Shaanxi 710049, China}

\author{Yihao Zhou}%
\affiliation{School of Physics, Xi'an Jiaotong University, Xi'an, Shaanxi 710049, China}

\date{\today}%

\begin{abstract}
Utilizing quantum coherence monotone, we reexamine the thermal nature of the Unruh effect of an accelerating detector. We consider an UDW detector coupling to a $n$-dimensional conformal field in Minkowski spacetime, whose response spectrum generally exhibits an intermediate statistics of $(1+1)$ anyon field. We find that the thermal nature of the Unruh effect guaranteed by KMS condition is characterized by a vanishing asymptotic quantum coherence. We show that the time-evolution of coherence monotone can distinguish the different thermalizing ways of the detector, which depends on the scaling dimension of the conformal primary field. In particular, for the conformal background with certain scaling dimension, we demonstrate that at fixed proper time a revival of coherence can occur even for growing Unruh decoherence. Finally, we show that coherence monotone has distinct dynamics under the Unruh decoherence and a thermal bath for a static observer.
\end{abstract}

\maketitle

\section{Introduction}
\label{1}

The celebrated Unruh effect manifests the observer-dependence of the particle content of quantum fields, which happens in various scenarios of quantum gravity. It predicts \cite{UNRUH1} that an observer with uniform acceleration $a$ would perceive the ordinary Minkowski vacuum as a thermal state at temperature $T_U=a/2\pi$. The oft-heard interpretation of the Unruh effect  is relied on so-called thermalization theorem \cite{THERMAL1,THERMAL2} which claims that for an accelerating observer, the field degrees screened behind a causally disconnected Rindler wedge should be traced over, and leads to an information loss presented by a Gibbs state with temperature $T_U$. In this light, the Unruh effect can be featured by the entanglement entropy of the improper mixed modes separated by the Rindler horizon. In an axiomatic way \cite{UNRUH6,UNRUH7}, the Unruh effect is a direct consequence of the Bisognano-Wichmann theorem \cite{UNRUH8}, then is universal in arbitrary dimensions independently of whether or not particles are interacting. Beyond these formal studies,  more realistic way to manifest the thermal nature of Unruh effect is to utilizing an Unruh-DeWitt (UDW) detector \cite{UNRUH5}. Once for instance coupling to a massless field in Minkowski spacetime, the detector following an accelerating trajectory would be excited and exhibit a power spectrum in Planckian form, agreeing precisely with the conventional thermal spectrum for a static trajectory in a heat bath. The parameter $T_U$ appearing in detector's spectrum is then recognized as a temperature. 

However, there are some subtle differences between the Unruh radiation and a thermal bath in static frame. Indeed, for an accelerating UDW detector, its Planckian excitation spectrum may merely be a coincidence, which is broken once the detector couples to a massive background field \cite{TAKAGI2} or under a spinning motion \cite{TAKAGI3}. It was soon realized that the thermalization theorem behind the Unruh effect only guarantees the Kubo-Martin-Schwinger (KMS) condition, or equivalently the principle of detailed balance, and nothing more \cite{TAKAGI1}. The insight is further convinced through the notable phenomenon called \emph{statistics inversion} \cite{TAKAGI4}, which indicates that the statistics endowed from detector's \emph{response spectrum} sensitively depends on the spacetime dimensions $n$. For a massless scalar background, the Bose-Einstein distribution of detector's excitation for even $n$ would be replaced by a Fermi-Dirac one for odd $n$, even without any fermionic field degrees. This clearly not the case for a static detector immersed in a thermal bath, whose response spectrum maintains Planckian for arbitrary spacetime dimensions. However, the statistics inversion does not spoil the KMS condition from which the effective temperature $T_U$ can be recognized \cite{TAKAGI5}. The formal reason behind the anomalous statistics was found \cite{TAKAGI6} that in odd dimensional spacetime the particular analytic structure of two-point Wightman function leads to the anticommutativity for scalar correlator and breaks the Huygens principle hold in even dimensional spacetime. It was further clarified in \cite{TAKAGI7,TAKAGI8} that the statistics inversion should be viewed as a {local} feature of the detector, encoding the nontrivial density of states on the Rindler wedge, or the information about the way of the detector's state to approach thermal equilibrium. 

Essentially, the Unruh radiation is a quantum feature in relativistic frame. Recently, there is growing interests in further exploiting the Unruh effect by some probes of quantumness, beyond the response spectrum of UDW detector. The inviting candidates include the entanglement monotone between multi-detectors \cite{OP2,OP3,OP4}, quantum discord \cite{OP+1}, uncertainty relation \cite{OP5,OP6,OP7,OP8} and quantum Fisher information \cite{OP9,OP10,OP11}. Besides revealing the unique quantum feature that absence for conventional thermal bath noise, these quantum probes may also shed new light in quantum gravity \cite{OP12}.

In this paper, we will exploit the quantum coherence (QC) \cite{coh-1,coh-2,coh-3}, another general character of quantumness, for an accelerating UDW detector coupling to a background field of interested. Comparing to the aforementioned studies \cite{OP11,OP11+1,OP11+2}, two generalizations made in present paper worth to be highlighted. Firstly, beyond quantum correlation of multipartite system, QC encapsulates the defining features of quantum theory in arbitrary dimensions, including superposition principle of single qubit. It constitutes a powerful physical resource that exhibit advantage in certain metrological task where entanglement does not \cite{OP11}. Secondly, extending usual focusing on Bose or Fermi statistics of local response, we study an UDW detector coupling to a conformal scalar background field, which leads to a detector's response spectrum of intermediate statistics \cite{TAKAGI10}, and includes Takagi's statistics inversion as a special case \cite{TAKAGI11}. 

We will show that the universal thermal nature of Unruh effect can be represented by the asymptotically vanishing QC monotone, guaranteed by the KMS condition from the thermalization theorem. On the other hand, the different thermalizing ways of the detector, depends on the scaling dimension of conformal primary field operator, can be distinguished by the time-evolution of detector's coherence. We will show that for a particular range of scaling dimension, the QC of single UDW detector can be amplified and exhibit a revival with growing Unruh temperature. Finally, we will demonstrate that like other quantum probes, QC dynamics under the Unruh decoherence is distinct from the one related to a conventional thermal bath in a static frame. We will derive our results by treating the UDW detector as an open quantum system whose dynamics is governed by a Lindblad-type master equation \cite{OP1}, thus its Markovian evolution can be resolved for times much longer than the thermalization time. Throughout the paper, we admit the natural units as $\hbar=c=k_B=1$.


\section{Detector within a conformal background}
\label{2}


\subsection{The set-up} 
Following the approach of \cite{OP2}, we treat the accelerating UDW detector as an open quantum system coupled to a conformal background field $\mathcal{O}_\Delta$ in $n$-dimensional Minkowski spacetime. We consider the combined system of detector and quantum field with its total Hamiltonian given by
\be
H_{\text{total}}=H_{\text{detector}}+H_{\mathcal{O}_\Delta}+\mu H_{\text{int}}             \label{eq1}
\ee
where UDW detector is modeled by a two-level atom $H_{\text{detector}}=\frac{1}{2}\omega \sigma_3$, coupling to the conformal field $\mathcal{O}_\Delta$ through $H_I=(\sigma_++\sigma_-)\mathcal{O}_\Delta$. Here, we denote the atomic raising and lowering operators as $\sigma_\pm$, and $\omega$ is the energy level spacing of the atom. In $n$-dimensional Minkowski spacetime, the primary operator $\mathcal{O}_\Delta$ is characterized by a two-point Wightman function $G^{+}_{\Delta}\left(x, y\right)\equiv\left\langle 0\left|\mathcal{O}_{\Delta}(x) \mathcal{O}_{\Delta}(y)\right| 0\right\rangle$ respecting its conformal symmetry \cite{CFT}
\be
G^{+}_{\Delta}\left(x, y\right)=\frac{\Gamma(2 \Delta) {e}^{-i \operatorname{sgn}\left(x^{0}-y^{0}\right) \pi \Delta}}{2 \pi \left(x-y\right)^{2\Delta}}          \label{eq2}
\ee
for a timelike separation $(x-y)^2<0$. Here the scaling dimension $\Delta$ is determined under a dilatation transformation like $x\rightarrow \lambda x$, which provides an intrinsic {positive} parameter of the field \cite{CFT1}.

The state of combined system $\rho_{tot}$ unitarily evolves and is govern by von Neumann equation $\dot{\rho}_{tot}(\tau)=-i[H,\rho_{tot}(\tau)]$, where $\tau$ is the proper time of the detector. Assuming a weakly coupling between the detector and the field environment ($\mu\ll1$), the initial state of combined system can then be approximated as $\rho_{tot}(0)=\rho(0)\otimes |0\ra\la0|$, where $\rho(0)$ is the initial state of the atom and $|0\ra$ is the field vacuum respecting spacetime symmetry. The reduced dynamics of detector's state can then be derived by integrating over the background field degrees from $\rho_{tot}(\tau)$, represented by a Kossakowski-Lindblad master equation \cite{OP14,OP15}
\be
\dot{\rho}(\tau)=-i\left[H_{\mbox{\tiny eff}},\rho(\tau)\right]+\mathcal{L}\left[\rho(\tau)\right]             \label{eq2+}
\ee
where
\be
\mathcal{L}\left[\rho\right]=\sum^3_{i,j=1}C_{ij}\left[\sigma_j\rho \sigma_i-\frac{1}{2}\left\{\sigma_i\sigma_j,\rho\right\}\right]        \label{eq3}
\ee
The matrix $C_{ij}$ can be explicitly determined through a decomposition
\be
C_{ij}=\frac{\gamma_+}{2}\delta_{ij}-i\frac{\gamma_-}{2}\epsilon_{ijk} n_k+\gamma_0\delta_{3,i}\delta_{3,j}          \label{eq4}
\ee
where Kossakowski coefficients are
\be
\gamma_\pm= \mathcal{G}_{\Delta}(\omega)\pm \mathcal{G}_{\Delta}(-\omega),~~~\gamma_0=\mathcal{G}_{\Delta}(0)-\gamma_+/2         \label{eq5}
\ee
Here, the frequency Wightman function $\mathcal{G}_{\Delta}(\lambda)$ is obtained from the Fourier transform of (\ref{eq2}), i.e., $\mathcal{G}_{\Delta}(\lambda)=\int_{-\infty}^{\infty}d\tau~e^{i\lambda\tau}G^+_{\Delta}(\tau)$. Moreover, the interaction with external field would also induce a Lamb shift contribution for the detector effective Hamiltonian $H_{\mbox{\tiny eff}}=\frac{1}{2}\tilde{\omega}\sigma_3$, in terms of a renormalized frequency $\tilde{\omega}=\omega+i[\mathcal{K}(-\omega)-\mathcal{K}(\omega)]$, where $\mathcal{K}(\lambda)=\frac{1}{i\pi}\mbox{P}\int_{-\infty}^{\infty}d\omega\frac{\mathcal{G}_{\Delta}(\omega)}{\omega-\lambda}$ is Hilbert transform of Wightman functions. 

For the UDW detector modeled by a two-level atom, its density matrix can be expressed in a Bloch form
\be
\rho(\tau)=\frac{1}{2}\Big(1+\sum_{i=1}^3\rho_i(\tau)\sigma_i\Big)        \label{eq6}
\ee 
Substituting it into the master equation (\ref{eq2+}), the Bloch vector $\bm{\rho}=(\rho_1,\rho_2,\rho_3)^T$ satisfies \cite{OP2}
\be
\frac{\partial \bm{\rho}}{\partial \tau}+2 \bm{K}\cdot\bm{\rho}+\bm{\eta}=0 
\ee
where $\bm{\eta}=(0,0,-2\gamma_-)^T$ and $\bm{K}$ is
\be
\bm{K}=\left(\begin{array}{ccc}
 \gamma_++\gamma_0 & {\tilde{\omega}} / 2 & 0 \\
-{\tilde{\omega}} / 2 &  \gamma_++\gamma_0 & 0 \\
0 & 0 &  \gamma_+
\end{array}\right)      
\ee
Given a general initial state $|\psi\ra=\sin\frac{\theta}{2}|0\ra+\cos\frac{\theta}{2}|1\ra$, the master equation (\ref{eq2+}) can be analytically resolved. The time-dependent components of the state are given as
\bea
\rho_1(\tau)&=&e^{-\frac{1}{4}\gamma_+\tau}\sin\theta\cos\tilde{\omega}\tau\no\\
\rho_2(\tau)&=&e^{-\frac{1}{4}\gamma_+\tau}\sin\theta\sin\tilde{\omega}\tau\no\\
\rho_3(\tau)&=&e^{-\frac{1}{2}\gamma_+\tau}\left(\cos\theta+\gamma\right)-\gamma            \label{eq7}
\eea
where $\gamma\equiv\gamma_{-} / \gamma_{+}$ is a ratio of Kossakowski coefficients.


\subsection{Detector's state with intermediate statistics}
To derive the complete dynamics of detector's state (\ref{eq7}), we need to specify $\mathcal{G}_{\Delta}(\omega)$ following the trajectory of atom undergoing an uniformly acceleration, like
\be
x^0(\tau) =a^{-1} \sinh a \tau, ~ x^1(\tau)=a^{-1} \cosh a \tau ,~\bm{x}(\tau)=0           \label{eq8}
\ee
in $n$-dimensional Minkowski spacetime. 

Along the acceleration trajectory, the $n$-dimensional two-point Wightman function (\ref{eq2}) should be rewritten in the accelerating frame. By substituting (\ref{eq8}) into (\ref{eq2}), we obtain
\be
G^+_{\Delta}(\tau)=\frac{\Gamma(2 \Delta) }{2 \pi}\left[\frac{\pi T_U}{i \sinh \left(\pi \tau T_U\right)}\right]^{2 \Delta}         \label{eq9}
\ee
with Unruh temperature $T_U=a/2\pi$ is denoted. It is interesting to note \cite{TAKAGI10} that above Wightman function exactly coincides with the chiral part of the Wightman function of a $(1+1)$-dimensional anyon field at a same temperature $T_U$.

Under a Fourier transform, we obtain
\be
\mathcal{G}_{\Delta}(\omega)=\frac{(2 \pi T_U)^{2 \Delta-1} }{2 \pi}{e}^{\frac{\omega}{2 T_U}}\left|\Gamma\left(\Delta+\frac{i \omega}{2 \pi T_U}\right)\right|^{2}         \label{eq10}
\ee
which is a function of detector's energy spacing $\omega$, Unruh temperature $T_U$ and the scaling dimension $\Delta$ of background. 

We note that (\ref{eq10}) enjoys several important properties. Firstly, it fulfills 
\be
\mathcal{G}_{\Delta}(\omega)=e^{ \omega/T_U} \mathcal{G}_{\Delta}(-\omega)       \label{eq11}
\ee
which is a frequency version of Kubo-Martin-Schwinger (KMS) condition $G^{+}(\tau)=G^{+}(\tau+i /T_U)$ guaranteeing the thermal nature of Unruh effect \cite{TAKAGI1}. Secondly, using Euler reflection formula $|\Gamma(i x)|^{2}=\pi /(x \sinh (\pi x))$, one can rewrite (\ref{eq10}) in an instructive form for integer or a half-integer $\Delta$ as \cite{TAKAGI11}
\be
\mathcal{G}_{\Delta}(\omega)= \begin{cases}\displaystyle \frac{ \omega^{2 \Delta-1}}{1-{e}^{-\frac{\omega}{T_U}}} \prod_{k=0}^{\Delta-1}\left[1+\left(\frac{2 k \pi T_U}{\omega}\right)^{2}\right] & \Delta\in \mathbb{Z} \\ 
\displaystyle\frac{ -\omega^{2 \Delta-1}}{1+{e}^{-\frac{\omega}{T_U}}} \prod_{k=\frac{1}{2}}^{\Delta-1}\left[1+\left(\frac{2 k \pi T_U}{\omega}\right)^{2}\right] & \Delta\in \mathbb{Z}-\frac{1}{2}\end{cases}    \label{eq12}
\ee
which are polynomials of $\omega$ and involve Bose-Einstein and Fermi-Dirac factors. Once the scaling dimension \emph{happens to be} selected as $\Delta=(n-2) / 2$, these frequency Wightman functions just coincide with Takagi's statistics inversion for different spacetime dimensionality. That is the detector response exhibits a Planckian form in even dimensional spacetime, but reverse to a Fermi-Dirac distribution in odd dimension\footnote{For example, if taking $\Delta=1$, one can recover well-known response spectrum from (\ref{eq12}) for a massless scalar field in $n=4$ dimensional spacetime.}. In this light, for generic $\Delta$, (\ref{eq10}) provides a continuously interpolation between the Bose-Einstein and Fermi-Dirac statistics. Finally, we should emphasize that the frequency response (\ref{eq10}) in Unruh effect is essentially different from a thermal bath at a same temperature $T_U$ \cite{OP13}
\be
\mathcal{G}_{\text {thermal }} \sim-\frac{2^{2-n} \pi^{\frac{1-n}{2}} \omega^{n-3}}{\Gamma\left(\frac{n-1}{2}\right)} \frac{1}{e^{-\beta \omega}-1}  \label{eq12+}
\ee
which contains a Planckian factor for arbitrary spacetime dimensionality $n$. Thus the statistics inversion encoded in (\ref{eq12}), with $\Delta=(n-2) / 2$ has been chosen, is absence for a thermal bath in static frame.

Substituting (\ref{eq10}) into (\ref{eq5}), we obtain the related Kossakowski coefficients for the conformal background
\bea
\gamma_+&=&\frac{a^{2 \Delta-1} }{ \pi} \cosh \left(\frac{ \omega}{2T_U}\right)\left|\Gamma\left(\Delta+\frac{i \omega}{2 \pi T_U}\right)\right|^{2}\no\\
\gamma_-&=&\frac{a^{2 \Delta-1} }{ \pi} \sinh \left(\frac{ \omega}{2T_U}\right)\left|\Gamma\left(\Delta+\frac{i \omega}{2 \pi T_U}\right)\right|^{2} \no\\      
\gamma &=& \gamma_{-} / \gamma_{+}=\frac{1-e^{- \omega/T_U}}{1+e^{- \omega/T_U}}=\tanh ( \omega / 2T_U)       \label{eq13}
\eea
We note the ratio $\gamma$ depends on the Unruh temperature $T_{U}$ due to the frequency KMS condition (\ref{eq11}), but has nothing to do with local correlator of the background.


\section{QC of an accelerating UDW detector}


\subsection{QC monotone}
We first review some elementary concepts on coherence measures, and admit the same notion of coherence used in \cite{coh-1}. For a given fixed basis $\{|i\rangle\}$, the incoherent states $\mathcal{I}$ is the set of quantum states with diagonal density matrix with respect to this basis. Incoherent completely positive and trace preserving maps (ICPTP) are the maps that map every incoherent state to another incoherent state. 

Given above setting, we say that $\mathcal{C}$ is a proper measure of QC if following axioms are satisfied: 

(C1). $\mathcal{C}(\rho) \geqslant 0$ for any quantum state $\rho$. The inequality is saturated iff $\rho \in \mathcal{I}$. 

(C2a). The measure is non-increasing under a ICPTP map $\Phi$, i.e., $\mathcal{C}(\rho) \geqslant \mathcal{C}(\Phi(\rho))$. 

(C2b). Monotonicity for average coherence under selective outcomes of ICPTP: $\mathcal{C}(\rho) \geqslant$ $\sum_{n} p_{n} \mathcal{C}\left(\rho_{n}\right)$, where $\rho_{n}=K_{n} \rho K_{n}^{\dagger} / p_{n}$ and $p_{n}=\operatorname{Tr}\left[K_{n} \rho K_{n}^{\dagger}\right]$ for all $\{K_{n}\}$ with $\sum_{n} K_{n}^{\dagger} K_{n}=\mathbf{1}$ and $K_{n} \mathcal{I} K_{n}^{\dagger} \subseteq \mathcal{I}$. 

(C3). Convexity, i.e. $\lambda \mathcal{C}(\rho)+(1-\lambda) \mathcal{C}(\sigma) \geqslant \mathcal{C}(\lambda \rho+(1-\lambda) \sigma)$, for any density matrix $\rho$ and $\sigma$ with $0 \leqslant \lambda \leqslant 1$.

A general distance-based coherence quantifier can be found, satisfying all the conditions (C1)-(C3), in terms of the minimal distance between target state and a given incoherent state. Using quantum relative entropy
\be
S(\rho \| \sigma)=\operatorname{Tr}\left[\rho \log \rho\right]-\operatorname{Tr}\left[\rho \log \sigma\right] \label{eq14}
\ee
one coherence monotone is $\min _{\sigma \in \mathcal{I}} S(\rho \| \sigma)$ which can be further rewritten as \footnote{While alternative monotones like $l_1$-norm exist, we prefer (\ref{eq15+}) as the quantifier of coherence since it works even with infinite system and avoids potential divergent \cite{coh-4}. Moreover, in terms of quantum relative entropy, it also enables us to compare present results with future exploration on the nonequilibrium dynamics for the UDW detector system \cite{future}.}
\be
\mathcal{C}_{\text {R.E. }}({\rho})=S\left({\rho}_{\text {diag }}\right)-S({\rho})      \label{eq15+}
\ee
where $S(\hat{\rho})=-\operatorname{Tr}(\hat{\rho} \log \hat{\rho})$ is von Neumann entropy of the state $\hat{\rho}=\sum_{i j} \rho_{i j}\left|i\right\rangle\left\langle j\right|$, and $\hat{\rho}_{\text {diag }}=\sum_{i} \rho_{i i}\left| i\right\rangle\left\langle i\right|$ is derived via a dephasing operation on density matrix.


\subsection{QC with intermediate apparent statistics}

For a general single qubit (\ref{eq6}), its related QC monotone (\ref{eq15}) can be straightforwardly calculated as
\be
\mathcal{C}_{\text {R.E. }}({\rho})=H_{\text{binary}}\left(\frac{1+\rho_3(\tau)}{2}\right)-H_{\text{binary}}\left(\frac{1+\ell(\tau)}{2}\right)     \label{eq15}
\ee
where $\ell \equiv \sqrt{\sum_{i} \rho_{i}^{2}}$ is the length of Bloch vector and $H_{\text {binary }}(x) \equiv-x \log x-(1-x) \log (1-x)$ is the binary entropy of variable $x$.

Given the explicit density matrix of the UDW detector (\ref{eq7}) and (\ref{eq13}), it is easy to see the coherence monotone $\mathcal{C}_{\text {R.E. }}(\theta, \omega,T_U,\Delta;\tau)$ should be a time-depedent function of detector's energy gap $\omega$, initial state preparation $\theta$, the Unruh temperature $T_U$ as well as the scaling dimension $\Delta$ parameterizing the apparent statistics perceived by the detector. 

We know that after a sufficient long time, the detector should be thermalized. It is easy to see that for large time $\tau\rightarrow\infty$, the detector's state (\ref{eq7}) reaches at an equilibrium
\be
\rho_{\text{equil}}(T_U)=\frac{e^{- H_{\text{detector}}/T_U}}{\operatorname{Tr}\left[e^{- H_{\text{detector}}/T_U}\right]}           \label{eq16}
\ee
which is exactly a thermal state at Unruh temperature. Since (\ref{eq16}) is diagonal, (\ref{eq15}) becomes vanishing ascribed to the consistent Unruh decoherence. This indicates that the thermal nature of Unruh effect can be manifested by a vanishing asymptotic quantum coherence, i.e., $\mathcal{C}_{\text {R.E. }}(\tau\rightarrow\infty)=0$, which is irrelevant to the local nature of detector (e.g., initial state $\theta$ or the energy gap $\omega$) or the particularity of background field (scaling dimension $\Delta$).

\begin{figure}[hbtp]
\begin{center}
\hspace{-10pt}\subfloat[$\mathcal{C}_{\text {R.E. }}(\omega=1,T_U=1)$]{\includegraphics[width=.24\textwidth]{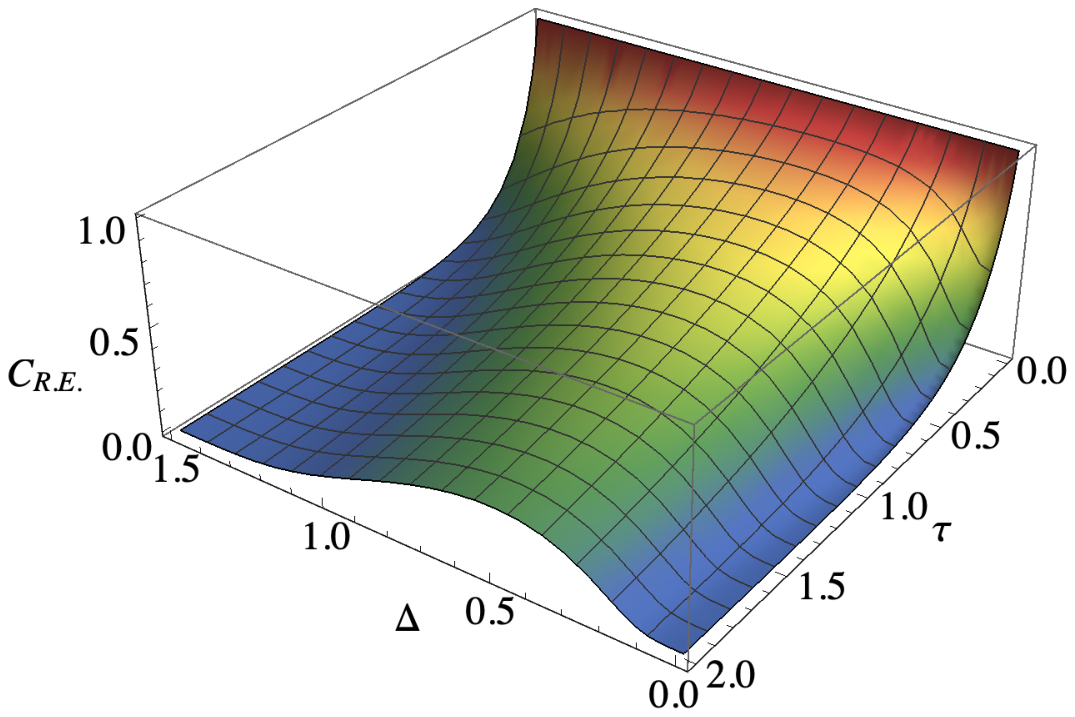}}~
\subfloat[$\mathcal{C}_{\text {R.E. }}(\omega=1,T_U=3)$]{\includegraphics[width=.24\textwidth]{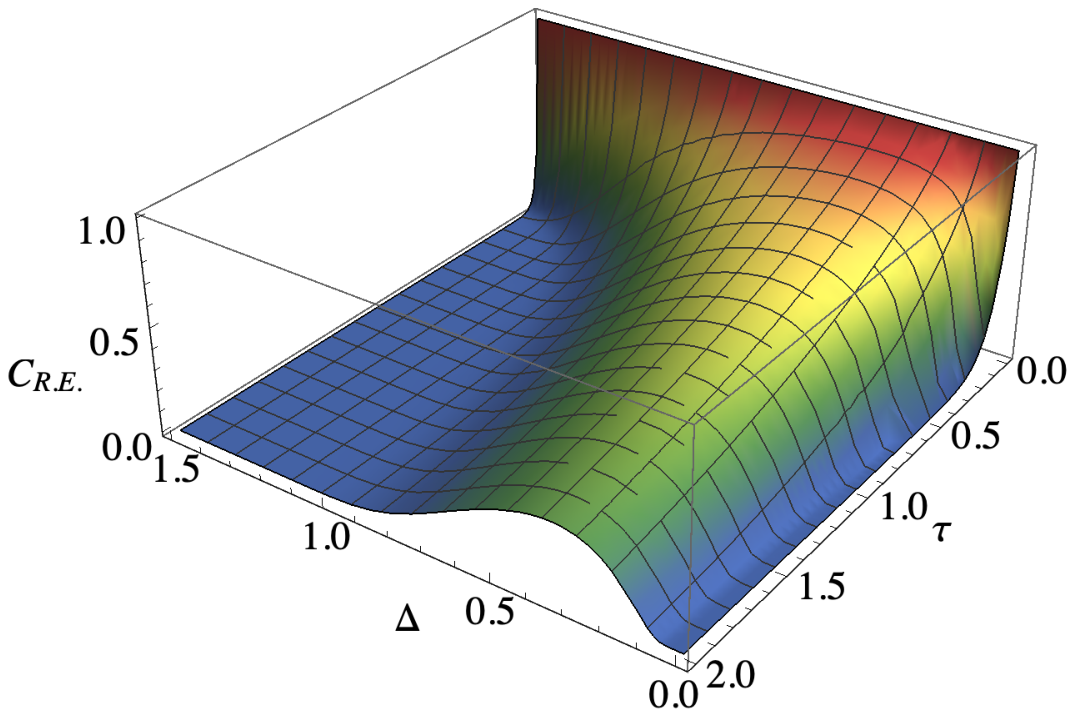}}
\caption{The quantum coherence $\mathcal{C}_{\text {R.E.}}$ of an accelerating UDW detector as a function of proper time $\tau$ and background scaling dimension $\Delta$. The detector's energy spacing and acceleration are taken as (a). $\omega=1,T_U=1$; (b). $\omega=1,T_U=3$, respectively.}
\label{fig1}
\end{center}
\end{figure}

The special way for detector's thermalizing to final equilibrium state (\ref{eq16}) can be recognized by investigating the time-evolution of QC monotone $\mathcal{C}_{\text {R.E.}}$. Noting that (\ref{eq15}) is periodic on $\theta$, without loss of generality, we will always fix $\theta=\pi(k+1)/2,~(k\in\mathbb{Z})$ thus maximizes the monotone. In Fig.\ref{fig1}, we depict the QC monotone $\mathcal{C}_{\text {R.E.}}$ as a function of detector's proper time and the scaling dimension of conformal background. We observe that the QC degrades monotonously with respect to the detector's proper time, until it approaching zero when the system thermalized to a final equilibrium. Nevertheless, with coupling to a conformal background with certain scaling dimension (particularly near $\Delta=0.5$), the related QC can be relatively \emph{robust} against to growing Unruh decoherence (e.g., see Fig.\ref{fig1}.(b) with a larger $T_U$). On the other hand, as aforementioned,  the detector's response spectrum with varying $\Delta$ may exhibit an intermediate statistics, which involves the statistics inversion in even/odd $n$-dimensional spacetime as a special case. However, no inversion behavior of $\mathcal{C}_{\text {R.E.}}$ as a function of scaling dimension (or spacetime dimensionality once adopting $\Delta=(n-2)/2$) was found. This was also the case in \cite{OP13} where another quantum probe, i.e., quantum Fisher information, was utilized. Thus we arrive our conclusion that the coherence evolution indeed distinguishes the way of the detector thermalizing, but is irrelevant with spurious interchange of statistics encoded in the functional form of response function \cite{TAKAGI8}.  

\begin{figure}[hbtp]
\begin{center}
\subfloat[$\mathcal{C}_{\text {R.E. }}(\omega=1,\tau=1)$]{\includegraphics[width=.21\textwidth]{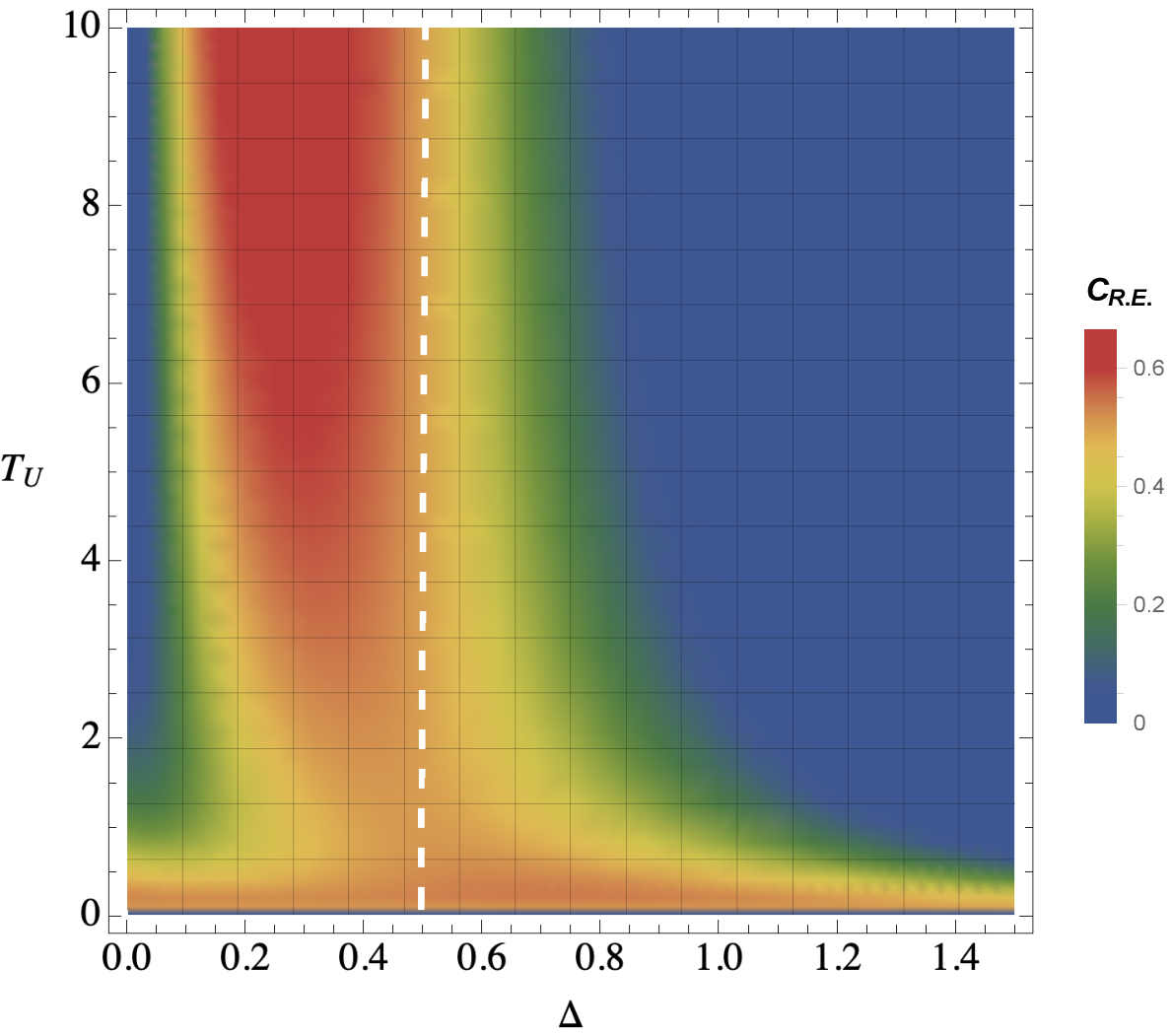}}
\subfloat[$\mathcal{C}_{\text {R.E. }}(\omega=1,\tau=1)$]{\includegraphics[width=.28\textwidth]{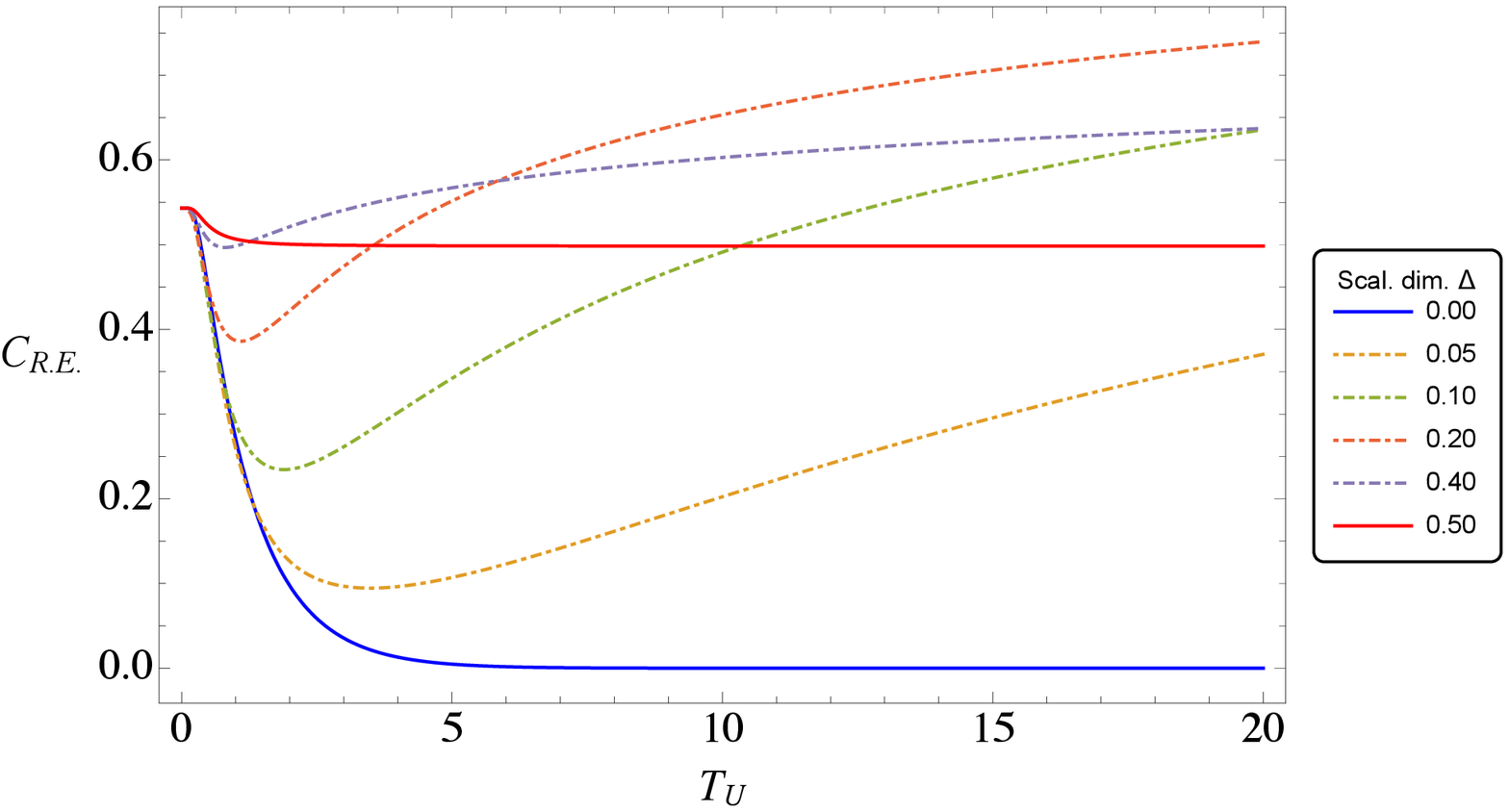}}
\caption{The quantum coherence $\mathcal{C}_{\text {R.E.}}$ as a function of the Unruh temperature $T_U$ and the scaling dimension $\Delta$ of the conformal background. The estimation is made at fixed time $\tau=1$ and $\omega=1$. In (a), the white dashed line indicates a critical value of $\Delta=0.5$, which bounds the left region wherein a revival of coherence can occur. In (b), the coherence revival is further demonstrated for particular choices of scaling dimension in the range $0<\Delta<1/2$.}
\label{fig2}
\end{center}
\end{figure}

Since the Unruh effect is usually recognized as an environmental decoherence, it is natural to expect that at fixed proper time, the QC monotone (\ref{eq15}) would also monotonously degrade with growing Unruh temperature. However, we find that it is subtle for a detector exhibiting an intermediate statistics. As plotted in Fig.\ref{fig2}.(a), with fixing $\tau=1$ and $\omega=1$, a revival behavior of coherence can occur, that is the coherence measure will reduce with growing $T_U$ firstly then rise up to an asymptotic value. As plotted in Fig.\ref{fig2}.(b), we further demonstrate this by plotting the coherence monotone as a function of Unruh temperature for a particular choice of $0\leqslant\Delta\leqslant 1/2$. It is shown that the coherence revival can only exhibit whence the conformal scaling dimension is restricted in the range $0<\Delta< 1/2$, otherwise the coherence monotone $\mathcal{C}_{\text {R.E.}}$ degrades monotonously with respect to $T_U$ as we expected.

\begin{figure}[hbtp]
\begin{center}
\hspace{-10pt}\subfloat[$\mathcal{C}_{\text {R.E. }}(\omega=1,\Delta=0.5)$]{\includegraphics[width=.24\textwidth]{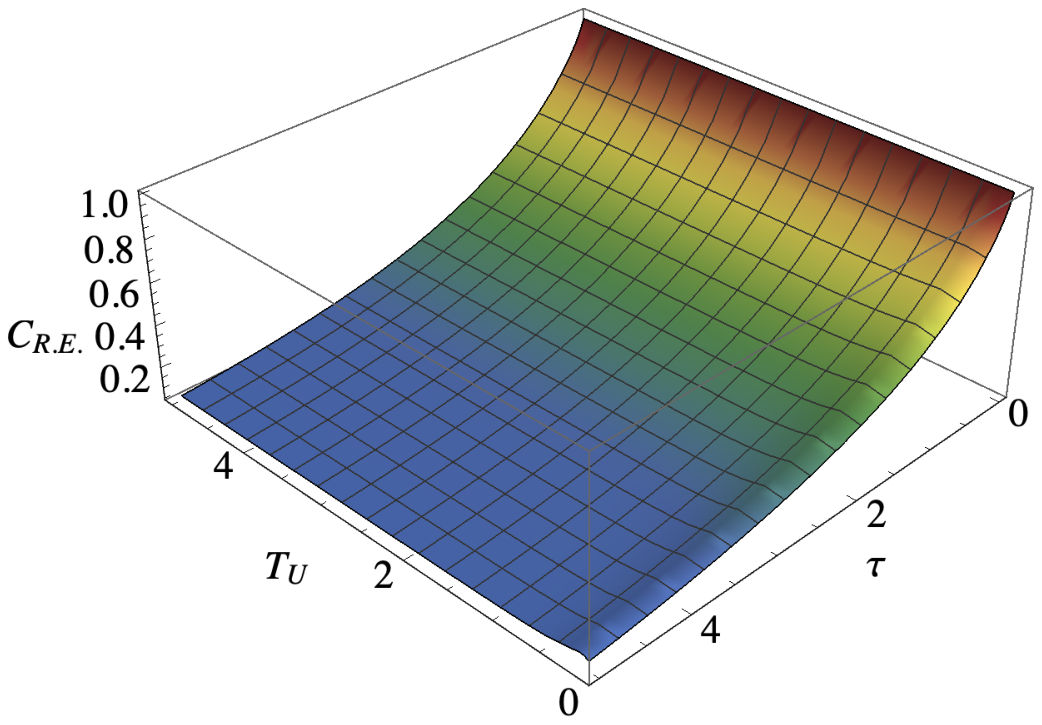}}~
\subfloat[$\mathcal{C}_{\text {R.E. }}(\omega=1,\Delta=1)$]{\includegraphics[width=.24\textwidth]{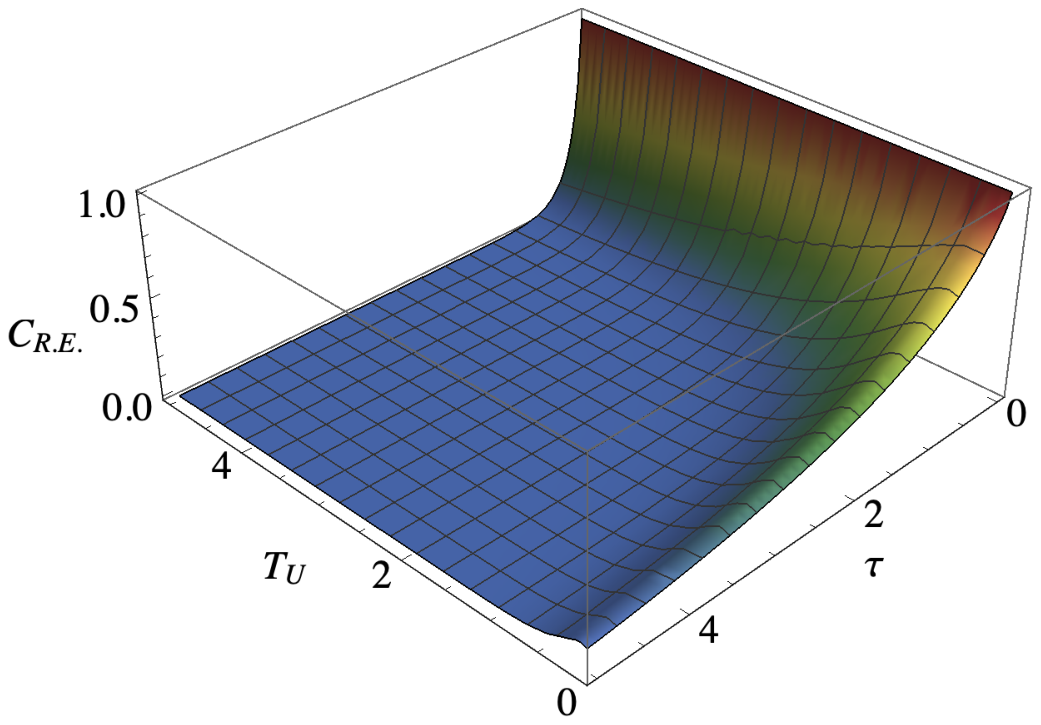}}\\
\hspace{-10pt}\subfloat[$\mathcal{C}_{\text {thermal }}(\omega=1,n=3)$]{\includegraphics[width=.24\textwidth]{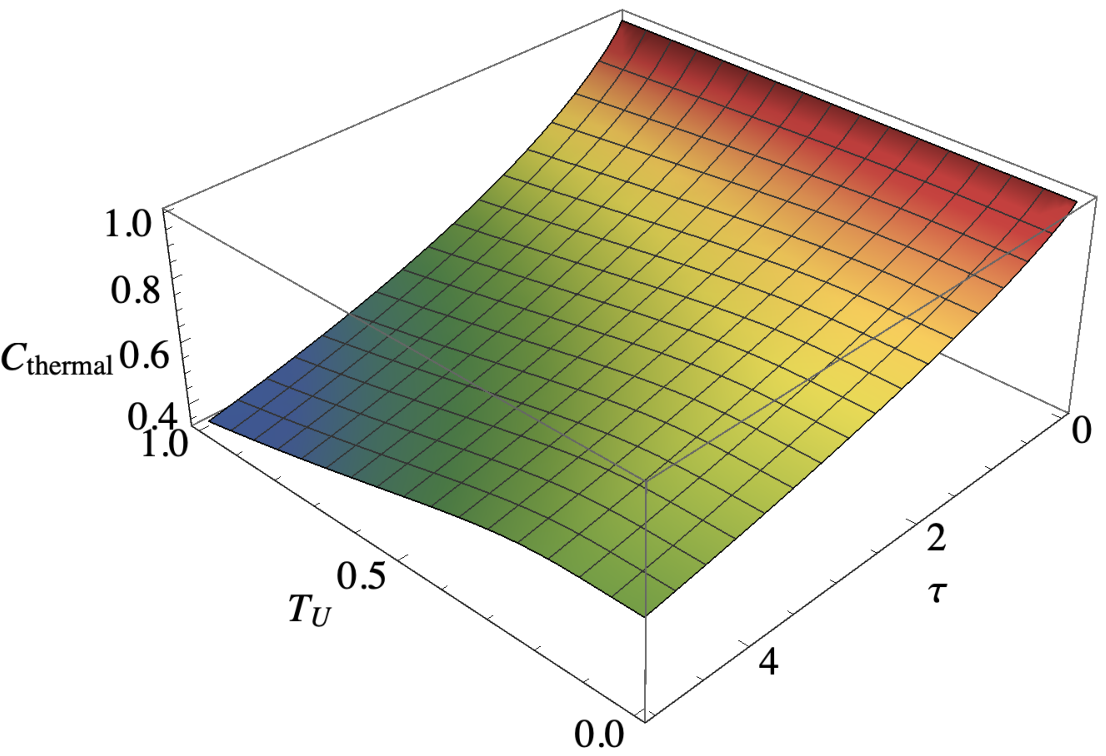}}~
\subfloat[$\mathcal{C}_{\text {thermal}}(\omega=1,n=4)$]{\includegraphics[width=.24\textwidth]{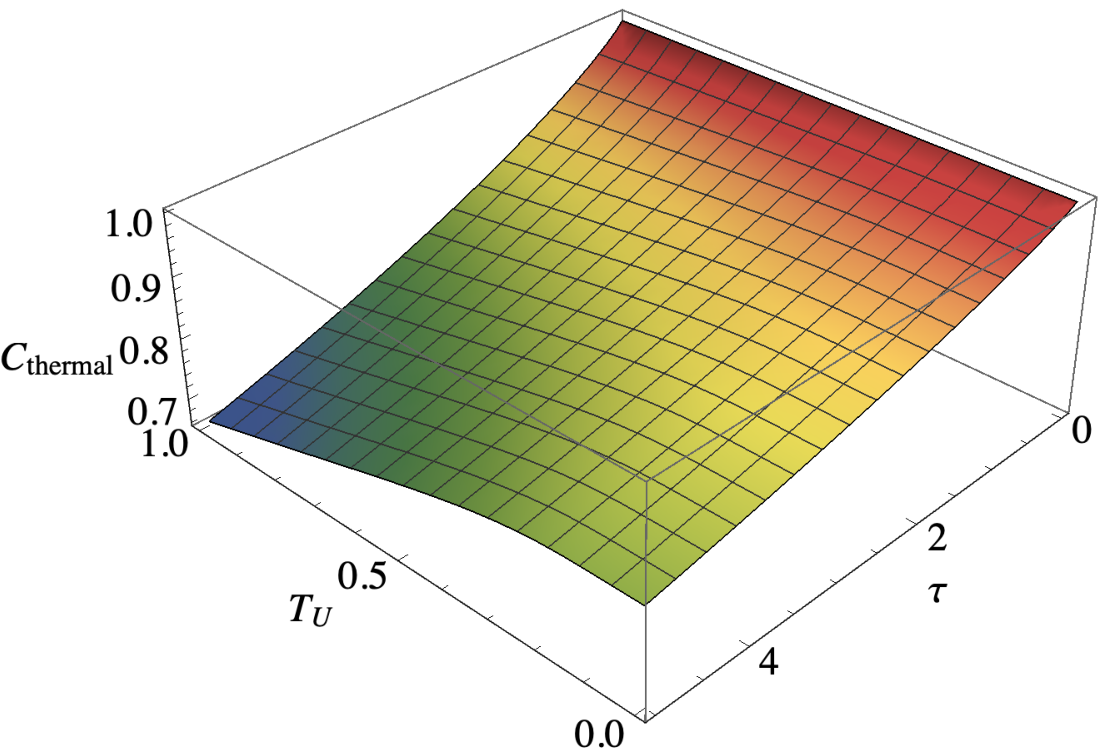}}
\caption{Quantum coherence distinguishes the Unruh effect and a conventional thermal bath, with the intrinsic scaling dimension has been chosen as $\Delta=(n-2)/2$, which gives $\Delta=0.5$ if $n=3$ and $\Delta=1$ for dimensionality $n=4$.}
\label{fig3}
\end{center}
\end{figure}

Finally, while the Unruh effect is essential different from a thermal bath perceived in static frame indicated from their response spectrums, we would like to show that such difference can also be recognized by comparing the QC of static detector in a thermal bath $\mathcal{C}_{\text {thermal}}$ with the Unruh case (\ref{eq15}). To arrive this, the Kossakowski coefficients related to a thermal response can be derived by substituting (\ref{eq12+}) into (\ref{eq5}). Then $\mathcal{C}_{\text {thermal}}$ are evaluated and plotted in Fig.\ref{fig3}.(c),(d), which are numerically different from the Unruh case shown in Fig.\ref{fig3}.(a),(b). We observe that the main difference between the Unruh and thermal QC is their distinct decay rates with respect to the proper time $\tau$, governed by the exponential factor $\sim \exp(-\gamma_+\tau)$ in density matrix (\ref{eq7}). For the Unruh effect, the particularity of the frequency response (\ref{eq12}) leads to a Kossakowski coefficient $\gamma_+$ modulated by a polynomial on the Unruh temperature $T_U$, whose power depends on the spacetime dimension \cite{OP13}. However, for a thermal background, such dimensionality dependence is absence since the response (\ref{eq12+}) maintains the same Planckian factor in arbitrary dimension. Therefore, as spacetime dimension varying, we are able to distinguish the Unruh decoherence from a thermal noise by the time-decaying QC of the detector.

\section{Summary and discussion}
\label{4}

In this paper, we investigate the thermal nature of the Unruh effect for an accelerating UDW detector by utilizing the coherence monotone as a quantum probe. Beyond the conventional scalar or fermionic background, we consider the detector that couples to a $n$-dimensional conformal background field, thus exhibits a spectrum with intermediate statistics coinciding with a $(1+1)$ anyon field. We show that the QC monotone of the detector $\mathcal{C}_{\text {R.E.}}$ has a vanishing asymptotic value for a sufficient long time, indicating the thermal nature of the Unruh effect guaranteed by KMS condition. Nevertheless, the different thermalizing paths of the detector, depending on the apparent statistics it perceived can be characterized by the time-evolution of coherence monotone. At fixed proper time, we find that the detector's QC may occur a revival phenomenon for the conformal background with $0<\Delta< 1/2$, that is it reduces firstly then rise up to an asymptotic value with growing Unruh temperature $T_U$. Note that coherence is by definition a basis-dependent quantity. If we adopt the viewpoint that a preferred basis of coherence only emerges by the environment \cite{Disc1}, with varying Unruh temperature, we would better cautiously refer the related background states as different environment. In this context, the coherence revival appearing for particular scaling dimension may be identified as a demonstration of ''einselection'' in relativistic framework. 

As a physical quantum resource, QC for multipartite system is interchangeable with other resources such as entanglement or discord \cite{coh-5}. It would be interesting to extend our analysis to multi-UDW detectors system and compare the QC dynamics with other quantum correlations. Indeed, some prior works have shown \cite{OP11} that the bipartite QC for two-UDW detectors within scalar background may also exhibit a QC revival that benefits some metrological tasks. By extending these arguments to the general conformal background presented in this paper, new light may shed on our interpretation for the quantum side of Unruh effect. The related work will be reported elsewhere.

\acknowledgments
This work is supported by the National Natural Science Foundation of China (No.12075178), the Fundamental Research Funds for the Central Universities, Natural Science Basic Research Plan in Shaanxi Province of China (No. 2018JM1049). J.F. would like to thank Yao-Zhong Zhang for enlightening discussions. He would also like to thank the kind hospitality at the University of Queensland where part of this work was accomplished.

\end{document}